\documentstyle[12pt,epsfig]{article}
\begin{document}
\begin{center}
\Large
{\bf The Scale Dependence of Inclusive {\it ep} Scattering
in the Resonance Region.}

\large
\vspace{1.7cm}
Simonetta Liuti$^{{\bf a},{\bf b}}$ \\
\vspace{0.5cm}
%
$^{\bf a}$ {\it Department of Physics, University of Virginia \\
McCormick Road. Charlottesville, Virginia 22901, USA. \\
\vspace{0.5cm}
$^{\bf b}$ INFN, Sezione di Roma Tre\\
Dipartimento
di Fisica E. Amaldi \\ Via Vasca Navale, 84. 00146 Roma, Italy. } 
\large
\vspace{1cm}

{\bf Abstract} \\
\end{center}
\vspace{-0.3cm}
\normalsize
We examine the scale dependence in the resonance region of inclusive 
{\it ep} scattering. In particular we discuss 
the  invariants other than $Q^2$,  
which have been proposed as a scale for pQCD 
evolution in a kinematical regime
where both infrared singularities and power corrections
are expected to be largest. We show that the 
region where most of the present data are available, can 
be described using NLO pQCD evolution at fixed invariant mass, 
$W^2$, plus a leading order power correction term. We find that 
the coefficient of the power correction at $W^2 < 4 \, GeV^2$ is 
relatively small, {\it i.e.} comparable in size to the one obtained in the 
large $W^2$ region.

\vspace{0.8cm}
\normalsize
\noindent
It has been long known that a 
fully quantitative description of proton structure 
in terms of parton distribution functions must account for
power corrections to the $Q^2$ 
dependence of the data, in addition to
the predicted perturbative-QCD (pQCD) behavior. 
Power corrections are indeed observed in experiments  
as discrepancies between fixed order perturbative predictions and the
data. Their theoretical interpretation is however  
a less well defined issue. 
In Deep Inelastic Scattering (DIS), in particular,
the short-distance scattering   
involving single, non-interacting partons 
is expected to give way to processes  
in which two (or more) quarks or gluons participate simultaneously in 
the scattering. These processes correspond formally to the 
Higher Twist (HT), or the higher order terms in the ``twist'' 
expansion (twist=dimension-spin) \cite{MirSan}. 
The coefficients of the HTs cannot be evaluated directly
within perturbation theory.   
However, it is also known 
that power corrections should appear in the coefficient functions
for hard processes, generated by the divergence
of the perturbative series at large orders (renormalons) \cite{Mue}.
Whether in DIS the two kinds of power corrections can be distinguished from 
one another and compared to the data is still an open question
(see however \cite{Katetal}).

In recent analyses the HT terms have been extracted from 
DIS data by applying 
a cut in the kinematics at $W^2 \geq 10 \, {\rm GeV^2}$ \cite{VirMil,Katetal}
($W^2$ is the invariant mass of the final hadronic state),
that is excluding the kinematical region dominated by nucleon 
resonances. 
Following \cite{DeRGeo,JiUnr} we have  
shown \cite{NicKep,EntKep} however that 
the entire set of inclusive data, including the 
low $W^2$ region can be described by a pQCD based analysis,  
the contribution from HTs being overall relatively small, {\it i.e.} 
within a factor of two from the one obtained 
in \cite{VirMil,Katetal}. 
The observation of a small power contribution can be otherwise phrased 
in terms of ``approximate Duality'', namely 
the non-perturbative features of the data 
appearing as the characteristic peaks describing the nucleon 
resonances, average out to a curve that  
can be identified with the DIS one modulo perturbative corrections
plus a small size power correction. 
Based on the results of \cite{NicKep,EntKep}, 
we perform here a more accurate analysis in the 
low $W^2$ domain with the aim of understanding 
the origin of the {\em residual} inverse-powerlike $Q^2$ dependence
of the data. 

Our analysis is based on three observations:

{\bf (1)} The recent Jefferson Lab data on the structure function $F_2$ \cite{NicPRL} 
show ``scaling'' in 
$W^2$, {\it i.e.} invariance with $W^2$ of the smooth curves which average
through the resonances peaks.
The smooth fits to the data plotted vs. $\xi = (2 x)/( 1 + 
+ 4 M^2 x^2 / Q^2 )^{1/2}$,
in order to account for target mass corrections, 
are shown in Fig.1 (dotted curves), for four different values of $W^2$ in the 
$W^2 \leq 4 \, {\rm GeV^2}$ range. 
Fig.1 also shows that as $W^2$ increases
some scaling violations are present.    
Eventually in the DIS region, {\it i.e.} at 
$W^2 \geq 4 \, {\rm GeV^2}$, $W^2$-scaling breaks down
completely.
\footnote{
This is due to the fact that the contribution of the  
fastly evolving sea quarks is no longer 
negligible.}
We conclude that the data {\em scale in $W^2$} 
so long as one one keeps inside the 
resonance region.

{\bf (2)} A pQCD based analysis is in principle possible in the resonance region
(see also \cite{DeRGeo,JiUnr,NicKep} so long as the values of $Q^2$ are 
larger than $\approx 1 \, {\rm GeV^2}$. Because of the kinematical
relation   
$
W^2 = Q^2 (1-x)/x + M^2,  
$
$M^2$ being the nucleon mass, this constraint corresponds to large 
$x$ values ($x \geq 0.2$ at present kinematics).
In analyses of DIS $F_2$ can be written as:
\begin{equation}
F_2(\xi,Q^2) = F_2^{NLO}(\xi,Q^2) \left(1 + \frac{C(\xi,Q^2)}{Q^2} \right),
\label{F2}
\end{equation}
where  $F_2^{NLO}(\xi,Q^2)$ is the NLO pQCD contribution 
and we disregard $O(1/Q^4)$ and higher terms. 
In order to try to reproduce $W^2$ scaling,
we consider evolution at fixed $W^2$. In other words in Eq.(\ref{F2}) 
$Q^2 = Q^2(x) \equiv (W_R^2 -M^2) x/(1-x) $ where $W_R$ is the fixed 
invariant mass of a given resonance. 
In our evolution equations we have used the recent PDFs from
\cite{GRV98}, in which $Q_o^2 \leq 1 \, {\rm GeV^2}$.
We limit our analysis to the large $x$ region ($x \geq 0.2$) 
so that the Singlet contribution that might introduce some ambiguity for 
the evolution down to a low scale, is negligible.  
The results of perturbative evolution, determining  $F_2^{NLO}$ are 
presented in Fig.1 along with the fits to Jlab data. 
We note first of all that there is an evident mismatch: 
the PDF results exceed the experimental values at $x<0.6$ and they lie
lower at large $x$. Secondly pQCD evolution predicts a stronger evolution at 
fixed $W^2$.   
  
Using Eq.(\ref{F2}) we interpret the discrepancies between perturbative
evolution and the data as given by  
the leading non-perturbative contribution
to the structure function. This actually enables us to determine 
the coefficient $C$ from the data:
\begin{equation}
C(\xi,Q^2) = \frac{Q^2 \Delta F}{F_2^{NLO}(x,Q^2)},
\end{equation}   
where
\begin{equation}
\Delta F  =  F_2^{NLO}(\xi,Q^2(x)) - F_2^{Exp}(x,Q^2(x))   
\end{equation}     
In Fig.2 we show $C(\xi,Q^2)$ vs. $\xi$.  
\begin{figure}[htb]
\vbox{
\hskip.6truecm\epsfig{figure=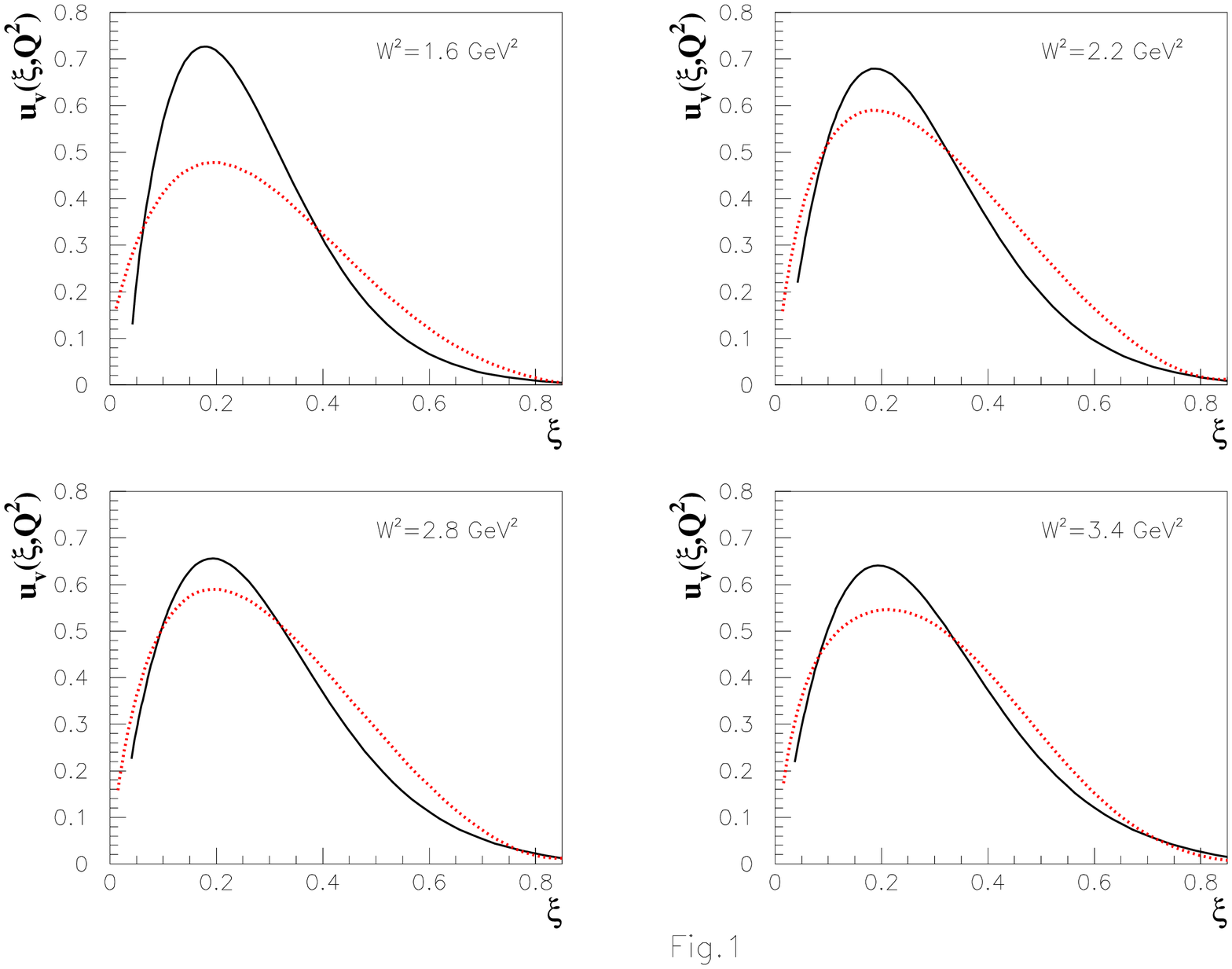,width=12.5truecm}}
\end{figure}
Our extractions are shown at different values of $Q^2$ (obtained by
transforming back from the fixed $W^2$ values). For comparison  
the results from the analysis of \cite{VirMil} using only $W^2 >10 \, {\rm GeV^2}$ data
are also shown (dotted curves).
\footnote{
Note: the appearence of a strong $Q^2$ dependence in the coefficient $C$ is mainly due
to the transformation $x \rightarrow \xi$.}  
Our results, which use only data at $W^2 \leq 4 \, {GeV^2}$, are in astonishing 
agreement with the high mass ones.   
At larger $\xi$ however the value of $C(\xi,Q^2)$ 
becomes less clearly determined and as
$\xi \rightarrow \xi_{th}$ (corresponding to $x \geq 0.8$) 
it is completely undefined.  
A possible interpretation of the  
indetermination at large $\xi$ is that standard DGLAP evolution
becomes unreliable and that evolution using a z-dependent scale 
should be performed instead \cite{Rob}. 
A quantitative approach to this problem 
is pursued in \cite{LiuEnt}. However, we also observe 
that at large $x$ data both in the large 
$Q^2$ region, determining $F_2^{NLO}$, and in the low
$W^2$ region, determining $F_2^{Exp}(x,Q^2(x))$ in our analysis, are
missing. Our analysis is extended to these regions by extrapolating what 
available at lower $x$ and its results are clearly less reliable here. 
These regions would be accessible at the $12$ GeV program at Jefferson
Lab.      

{\bf (3)} Finally, we comment on the interpretation of the power corrections in 
the resonance region. From a practical point of view, if power corrections are 
found to mantain the same $x$ dependence displayed in Fig.2,
namely $C \approx A/(1-x)$ as $x \rightarrow 1$, we predict that
\begin{equation}
R=
\frac{C(x,Q^2)}{Q^2(x)} \rightarrow \frac{A}{W_R^2-M^2},  
\; x \rightarrow 1
\end{equation}   
therefore at fixed $W^2$ one is approaching the $x \rightarrow 1$ limit and at 
the same time having control of the power correction terms. This situation is 
displayed in Fig.3 where we show R for different values of $W^2$.   
Not being dominated by power corrections, this region is ideal for 
pursuing further quantitative studies  
of deviations from NLO DGLAP evolution (\cite{PenRos} for earlier analyses
and {\it e.g.} \cite{Ste} for a more recent review).
\begin{figure}[htb]
\vbox{
\hskip.6truecm\epsfig{figure=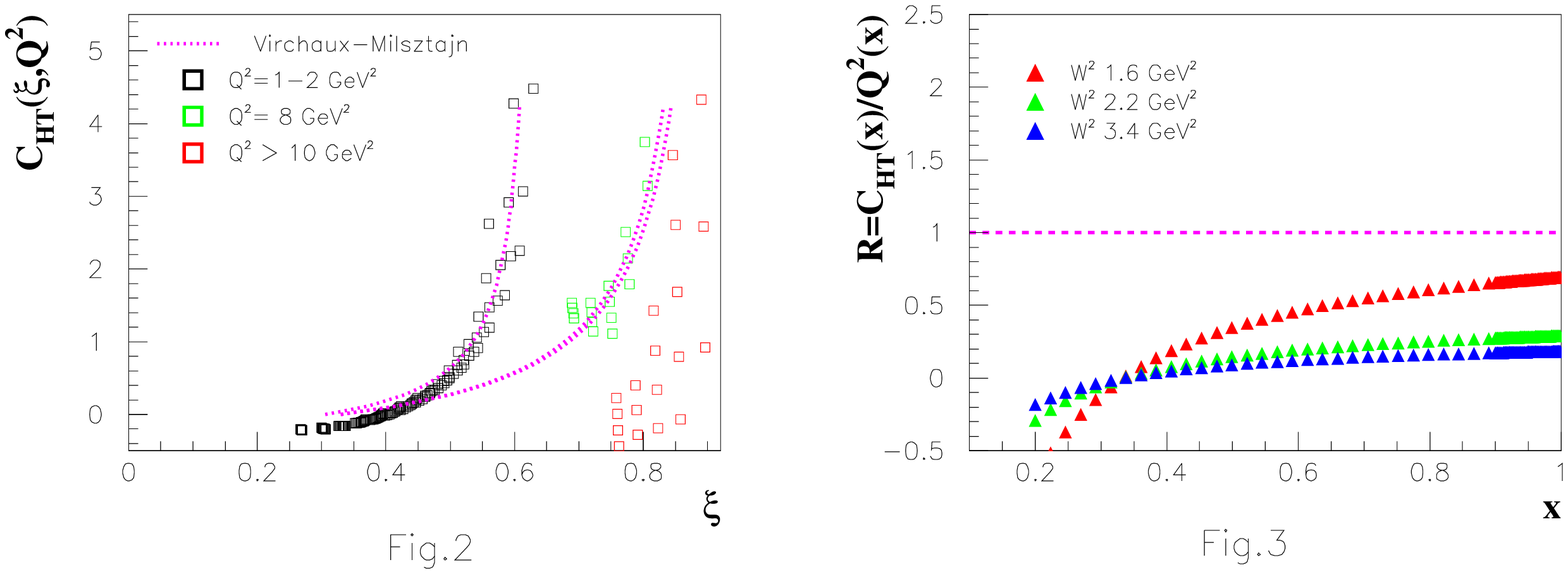,width=12.5truecm}}
\vskip-4.truecm
\end{figure}

What is the physical meaning of a small power correction in the resonance region? 
Power corrections might originate from the account 
of multi parton ``final state interaction'' processes that are more likely to occur 
at large distances and that correspond to well defined terms in the OPE. 
It is this type of interactions that are eventually responsible for 
confinement related features of the cross section such as the 
production of resonances. It turns out that for some, at present, unknown mechanism, 
the HTs contributions of increasing order in $1/Q^2$ cancel each other 
in the resonance
region, giving origin to the duality phenomenon 
(this is seen either in the average, Fig.1, or in the moments integrals \cite{DeRGeo}). 
However we find out through an accurate 
analysis of the data that duality is not exact: a ``residual'' $1/Q^2$ dependence with a 
coefficient comparable to the large $W^2$ analyses \cite{VirMil} is still necessary to
interpret the data. This $Q^2$ dependence not being ascribed to HT corrections, 
can be taken as the true contribution from the non-perturbative corrections to the
pQCD coefficient functions, namely the renormalon term. In order to confirm this 
interpretation, further studies addressed at determining the universality of this correction
should be performed. These would include studies of different structure functions, such
as $F_L$ in the resonance region, as well as scattering from different targets, 
including nuclei and studies of different fragmentation functions, a program accessible
at Jefferson Lab at $12 \, {\rm GeV}$.
 
On a more speculative level, the interpretation 
of the physical picture behind this behavior 
leads to a number of intriguing scenarios: for 
instance, partons inside hadrons might be arranged in a different way
at intermediate/large-distances, {\it e.g.} they might be clumped 
together inside valence quarks   
and the role of final state interactions might be 
effectively small down to low $Q^2$. 
  
In conclusion, information 
on the structure of the proton is still abundantly missing.
It could be obtained
if more data were available in the ``transition'' regions 
of $x$ and $Q^2$ where 
perturbative QCD (pQCD) evolution regulated by DGLAP equations 
can no longer be considered to be the main mechanism, 
and non-perturbative 
contributions become important.
  

\end{document}